# Assembling and Modeling Stacked Disordered Metasurfaces

Miao Chen [1*], Amit Sharma [2,3*], Johann Michler [2], Xavier Maeder [2], Philippe Lalanne [1], and Angelos Xomalis [2,4]

[1] LP2N, Institut d'Optique Graduate School, CNRS, Université de Bordeaux, Talence, 33400, France

[2] Laboratory for Mechanics of Materials and Nanostructures, Empa, Swiss Federal Laboratories for Materials Science and Technology, Thun 3602, Switzerland

[3] Swiss Cluster AG, Thun 3602, Switzerland

[4] Nanoelectronics and Photonics Group, Department of Electronic Systems, Norwegian University of Science and Technology, Trondheim 7034, Norway

**ABSTRACT**

Disordered metasurfaces offer unique properties unattainable with periodic or ordered metasurfaces, notably the absence of deterministic interference effects at specific wavelengths and angles. In this work, we introduce a lithography-free nanofabrication approach to realize cascaded disordered plasmonic metasurfaces with sub-micron total thickness. We experimentally characterize their angle-resolved specular and diffuse reflections using the bidirectional reflection distribution function (BRDF) and develop accurate theoretical models that remain valid even at large incidence angles. These models reveal the intricate interplay between coherent (specular) and incoherent (diffuse) scattering and demonstrate how coherent illumination can strongly influence the perceived color of diffusely scattered light. Exploiting this effect, we realize a centimeter-scale chromo-encryption device whose color changes depending on whether it is viewed under direct or diffuse illumination. Our results lay the groundwork for advanced nanophotonic platforms based on stacked disordered metasurfaces, offering versatile optical functionalities inaccessible with traditional multilayer thin-film technologies or single-layer metasurfaces.

**KEYWORDS**

Metasurfaces, disordered, cascading, thin-film, nanoparticles, encryption.

# INTRODUCTION

Optical metasurfaces, artificial two-dimensional arrangements of subwavelength scatterers, offer unprecedented control over the phase, polarization, and amplitude of electromagnetic waves [1,2]. This capability has spurred a wide array of applications, from structural color [3], sensors and displays [4,5], frequency conversion [6] and advanced imaging [7], to name a few. Within this domain, *disordered* metasurfaces—characterized by randomly positioned meta-atoms—present a compelling alternative [8,9]. While their theoretical analysis can be more complex than periodic counterparts [10,11], disordered metasurfaces offer significant advantages in cost-effective fabrication and scalability. Consequently, they have emerged as a versatile platform for large-scale applications, including sustainable renewable energy systems [12-14] and aesthetically vibrant coatings for luxury goods and architectural designs [15-23].

Similar to traditional optical systems, such as homogeneous film stacks and camera or microscope lenses, layering optical components offers a natural and potent strategy to enhance performance and unlock new functionalities, all while maintaining sub-micrometer thicknesses. This principle extends to Complex Optical Stacks (COSs), which are emerging as a cost-effective route to advanced optical functionalities by, for example, incorporating plasmonic nanoparticles into multilayer film architectures [24]. COSs hold significant potential for applications ranging from light extraction coatings and dynamic windows to touch screens and photovoltaics [25-28]. Within the realm of metasurfaces, multilayer stacks of *ordered* metasurfaces have already demonstrated success in broadening metalens operating bandwidths [29,30], correcting monochromatic aberrations [31], enabling complex wavefront shaping [32-34], and boosting harmonic generation [35]. However, the exploration of multilayer *disordered* metasurfaces—a specific and promising type of COSs—has remained comparatively limited, despite nature itself offering compelling examples [36]. Various species have evolved multilayered disordered structures that precisely shape both specular and diffuse reflected light, resulting in unique visual appearances, for example, the pointillist coloration of *Viburnum tinus* and *Pollia* fruits [37,38]. Inherent advantages of disordered structures are the suppression of deterministic interferential behavior, seen in periodic systems, and the preservation of stack planarity due to subwavelength particle dimensions, making cascaded disordered metasurfaces particularly attractive for robust and versatile optical control.

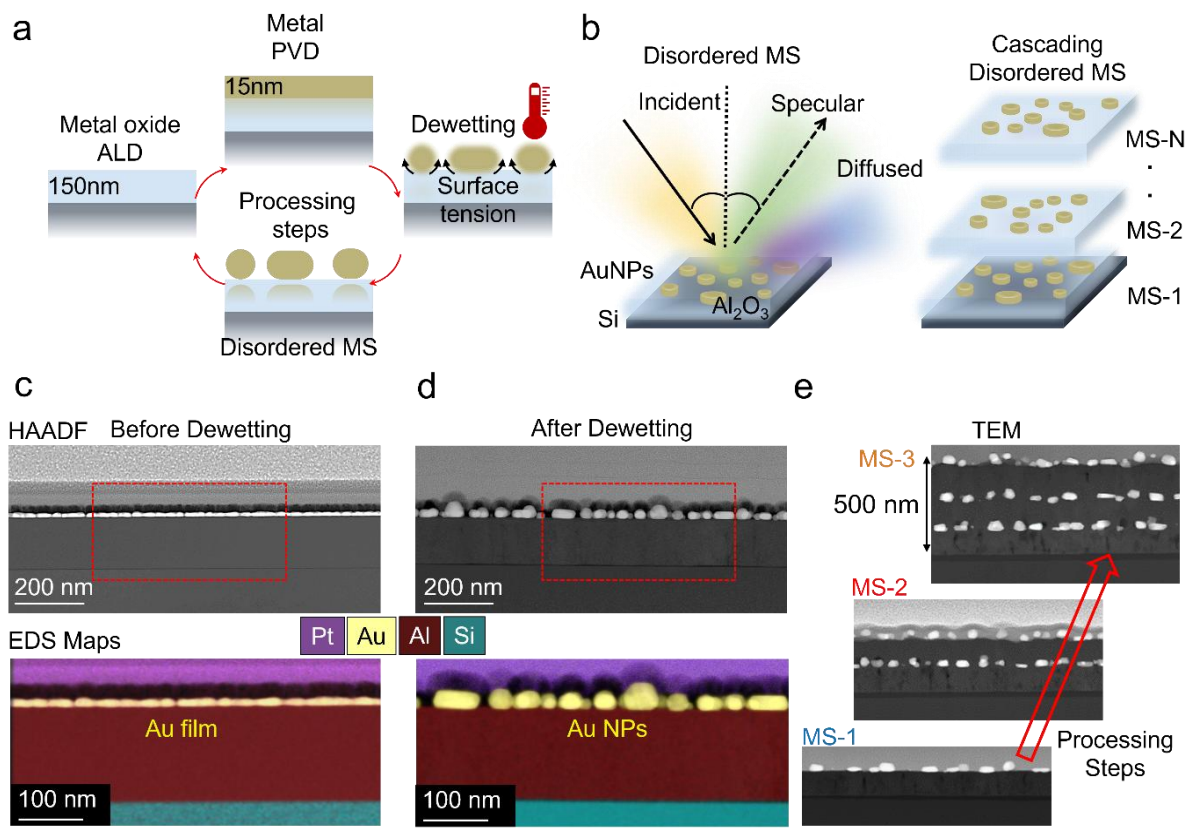

**Figure 1. Cascading disordered optical metasurfaces. a.** Assembling disordered metasurfaces via metal oxide atomic layer deposition (ALD) and metal sputtering within the same deposition reactor. Further large-scale dewetting on 4'' Si wafer. Surface tension induces the formation of Au nanoparticles (AuNPs). **b.** (Left) Unique visual appearance of disordered metasurfaces producing angularly separated specular (green) and diffused (violet) reflected light. (Right) Consecutive deposition and annealing resulting in cascading metasurface architectures. **c-d.** (Up) High-angle annular dark-field (HAADF) and (bottom) energy dispersive X-ray spectroscopy (EDS) transmission electron microscopy (TEM) images (left) before and (right) after dewetting. Here, we use $Al_2O_3$ as a dielectric spacer and 15 nm thick Au film deposited via ALD and sputtering, respectively. Inset: Colors stand for different

materials. **e** TEM images of the single- (MS-1), double- (MS-2) and triple-layered (MS-3) disordered MS. In all cases, the average size of AuNPs is about 50 nm.

Here we develop a general framework for the design and fabrication of miniaturized cascading disordered optical metasurfaces. Utilizing atomic layer deposition (ALD) with Angstrom thickness resolution for the dielectric spacer (~1.4 Å), metal sputtering, and large-scale dewetting of thin metallic films, we assemble plasmonic metasurface stacks, up to three-layered so far, with sub-micron (~500 nm) total thickness. To understand their optical behavior, we develop two complementary analytical models to predict both their specular and diffuse light responses of cascading metasurfaces. We further demonstrate that the stacking of metasurface layers modifies the coherent field exciting each individual layer, providing a mechanism to precisely shape both specular and diffuse colors. Leveraging this principle, we show how these designed disordered metasurface stacks can function as chromo-encryption surfaces, encoding information within their diffuse color. We anticipate that this integrated fabrication and simulation framework will facilitate the design of multilayer stacks of disordered metasurfaces, paving the way for the discovery and exploitation of more exotic optical phenomena in disordered photonic systems.

## RESULTS AND DISCUSSION

**Fabrication.**

To assemble cascading metasurfaces, we follow standard wafer-scale thin-film coating methodologies on silicon (Si). In Fig. 1, we illustrate the procedure for sample fabrication. With ALD and sputtering, we deposit metal oxide spacer ($Al_2O_3$) and 15 nm metallic film (Au), respectively (see Methods). Annealing induces dewetting of Au film that forms a metasurface comprising gold nanoparticles (AuNPs) of an average size of about 50 nm (Fig. 1c-d). By repeating deposition and annealing, we assemble ultrathin (sub-micron) disordered MS stacks. For obtaining nanostructural characterization, we use high-angle annular dark-field (HAADF) and energy dispersive x-ray spectroscopy (EDS) transmission electron microscopy (TEM) before and after dewetting (Fig. 1c-e). The ALD deposition of dielectric spacer produces well-spaced metasurfaces with thickness accuracy (1.4 Å, defined by the ALD growth per cycle, GPC). By fine tuning of the metal film thickness, we control the size and density of AuNPs [39].

The light scattered by disordered metasurfaces can be decomposed into specular and diffuse components [9]. The specular component is the statistically averaged electromagnetic field scattered by the metasurfaces. It is observed in the specular directions. The diffuse component represents the field fluctuations around the average.

**Model for the specular light component.**

To predict the specular optical response of the cascaded metasurfaces, we develop a semi-analytical model using multiple scattering theory. First, the reflection and transmission coefficients of each metasurface monolayer is modeled by assuming that the metasurface is buried in a homogeneous background medium with a refractive index equal to that of the metasurface medium in between the inclusions — air ($n = 1$) for the top MS, and $Al_2O_3$ ($n = 1.66$) for the other metasurfaces (Fig. 2a and 3). The computation of the reflection and transmission coefficients requires solving the electromagnetic scattering of a single inclusion (modeled as nanodisc) embedded in the homogeneous background [40,41] in combination with a freeware implementing near-to-far-field transformation [43]. Second, these coefficients are incorporated in a standard 2×2 transfer-matrix method [42] which additionally account for the propagation through the homogeneous 190nm-thick $Al_2O_3$ layers. Additional implementation details can be found in Suppl. Note 1 of the Supporting Information (SI). Input parameters for the metasurface stack, such as particle size or density, are extracted from scanning electron microscope (SEM) images of the sample. The refractive index of Au was taken from tabulated experimental data [44].

Fig. 2b provides a comparison between the theoretical (dashed lines) and experimental (solid lines) specular reflection spectra for single (MS-1), double (MS-2), and triple (MS-3) layered metasurfaces, for a planewave illumination with unpolarized light at an incidence angle $\theta_i = 10°$. The model reliably replicates the key experimental features, including a broad reflection peak around 700 nm with an intensity reaching approximately 40%. The addition of subsequent layers results in a slight narrowing of this peak but does not significantly shift its central wavelength or maximum intensity.

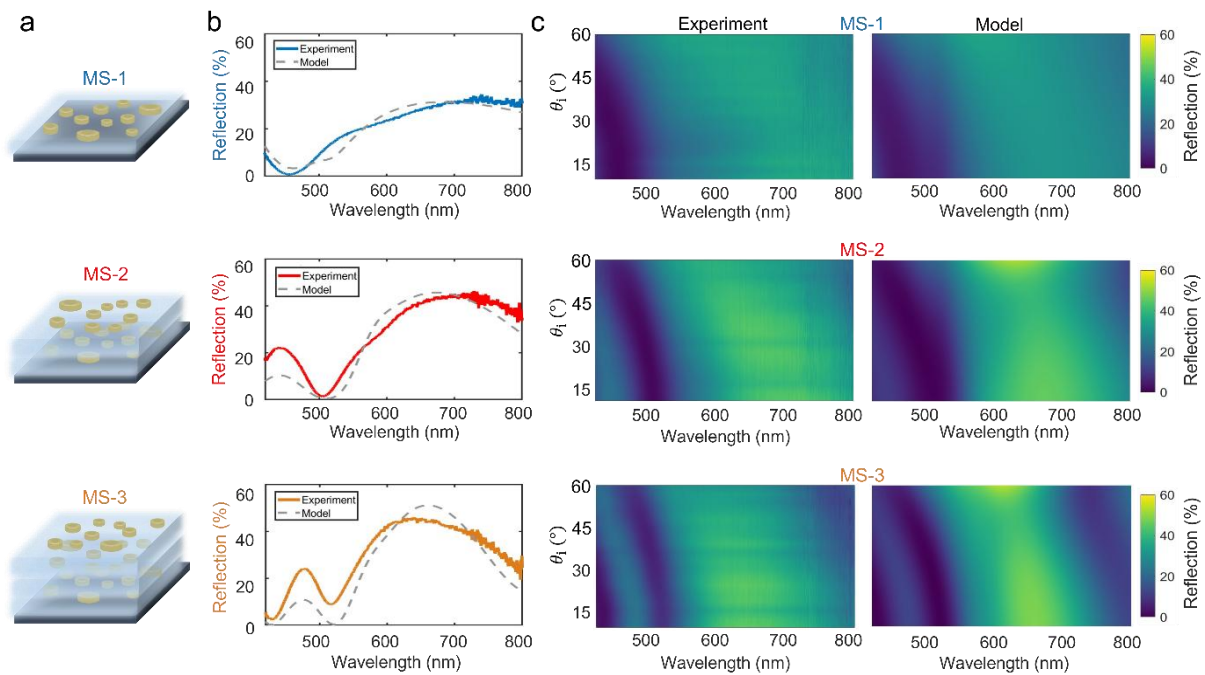

**Figure 2. Specular (coherent) reflectance of cascading metasurfaces. a.** Schematic representation of the metasurface stacks on a Si wafer and $Al_2O_3$ spacer layers of 190 nm. The average size of AuNPs is documented in Suppl. Note 3 in SI. **b.** Experimental specular reflection spectrum of a single (solid blue line, MS-1), double (red, MS-2) and triple (orange, MS-3) layered metasurfaces measured with a supercontinuum laser compared to modelled spectra (dashed grey line) at an incident angle $\theta_i = 10°$. **c.** Experimental (left) and modelled (right) angular-resolved specular reflection maps of MS-1 (top), MS-2 (middle) and MS-3 (bottom).

The predictive capability of the model was further evaluated using angle-resolved reflection measurements (Fig. 2c). The experimental maps (left column) show strong agreement with the model predictions (right column) across incident angles up to $\theta_i = 60°$. This agreement highlights the model effectiveness in capturing the angle-dependent multiple scattering phenomena inherent in these disordered systems, particularly at oblique incidence where such effects become more pronounced and rudimentary models often fail [9]. All results correspond to unpolarized illumination, with measurements taken using a setup utilizing a supercontinuum laser [22] and model calculations averaged over TE and TM polarizations (see Suppl. Note 1). The diameters used for these AuNP models corresponded to the average sizes extracted from SEM images (documented in Suppl. Note 3): 60 nm for MS-1, 52 nm for MS-2, and 42 nm for MS-3. The height of all nanodiscs is fixed at 35 nm. Thanks to its simplicity and analytical nature, the model allows for rapid computation of angle-resolved specular reflection maps within seconds.

**Model for the diffuse light component.**

The analysis of the diffuse light component scattered by the metasurface stack is performed by comparing theoretical predictions of the bidirectional reflection distribution function (BRDF) with experimental measurements. The BRDF is a multidimensional radiometric function, originally introduced in the 1960s, which essentially represents the angle-resolved far-field statistically average speckle intensity [9]. It characterizes how the diffused light is scattered into all directions for all incoming planewave illuminations.

Predicting the diffuse component of the BRDF of disordered media requires solving for the field-field correlation function, which, in the formalism of many-body theory, is represented by a vertex function [45]. This task is significantly more complex than computing the mean field that governs the specular response, making the study of diffuse light both theoretically and computationally demanding [9,45,46]. Approximate models are therefore necessary.

State-of-the-art analytical and numerical models for the diffuse BRDF of metasurface layers rely on the independent scattering approximation [9,22,23]. Valid primarily for low to moderate particle densities (typically $\rho <$ 1-2 µm$^{-2}$), the independent scattering approximation yields a particularly intuitive outcome: it decouples the contributions of individual scatterers—captured by the form factor—from those arising from their spatial arrangement—captured by the structure factor. This decomposition provides a powerful framework for understanding and engineering diffuse scattering.

Modern models relying on the independent scattering approximation include correction factors that approximately account for multiple scattering effects [9]. In this work, we extend this modeling approach to disordered metasurface stacks.

A seemingly straightforward, yet ultimately insufficient, strategy would be to treat the entire stack as a single disordered medium with a global structure factor. However, this simplification overlooks the depth-dependent variation ($z$-dependence) of both the incident and scattered fields across the stack vertical extent. In practice, particles in different layers experience distinct local excitations and contribute differently to the far-field scattering.

Further we present a simplified model for diffuse light scattering from metasurface stacks (see Suppl. Note S2). Under the assumptions that (1) the meta-atoms scatter independently and (2) the positions of the meta-atoms on one metasurface are statistically independent of those on the others, we show that the average speckle intensity of the stack is the incoherent sum of the average speckle intensities from each metasurface. This implies that the diffuse component $f_{\text{diff}}$ of the stack BRDF, i.e. the angle-resolved far-field statistically average speckle intensity scattered by the stack, can be expressed as a sum of the individual BRDF,

$$f_{\text{diff}} = \sum_{m=1,2,\dots} f_{\text{diff}}^{(m)}, \tag{1}$$

where $f_{\text{diff}}^{(m)}$ is the diffuse component of the BRDF of the metasurface labelled by the superscript ($m$).

In our simulations, we strictly follow the model in [9], which provides $f_{\text{diff}}^{(m)}$ for a single metasurface embedded in a thin-film stack (see Eq. 17 therein)

$$f_{\text{diff}}^{(m)} = \rho^{(m)} F^{(m)}(\mathbf{k}_{s,\|}, \mathbf{k}_{i,\|}, \hat{\mathbf{e}}_i) S_{r,\infty}^{(m)}(\mathbf{k}_{s,\|} - \mathbf{k}_{i,\|}) \frac{C^{(m)}(\mathbf{k}_{s,\|}, \mathbf{k}_{i,\|}, \hat{\mathbf{e}}_i)}{\cos\theta_i \cos\theta_s}, \quad (2)$$

where $\rho^{(m)}$ is the meta-atom density, $F^{(m)}$ is the meta-atom form factor and $S_{r,\infty}^{(m)}$ is the 2D structure factor of the metasurface $(m)$. $\mathbf{k}_{s,\|}$ and $\mathbf{k}_{i,\|}$ denote the in-plane wavevectors of the scattered and incident planewaves, and $\theta_s$ the polar angle of the planewave scattered in air. $C^{(m)}$ is a correction factor that approximately accounts for multiple scattering [22,47]. In reference [9], a label '+' or '−' is used in $F^{(m)}$ to distinguish between reflectance and transmittance. Since our analysis focuses solely on reflectance, this label is omitted here. Similarly, we ignore polarization effect and drop the polarization state $\hat{\mathbf{e}}_s$ of the scattered planewave in [9].

The z-dependence of the scattered field is incorporated by assigning a distinct form factor to each metasurface layer

$$F^{(m)}(\mathbf{k}_{s,\|}, \mathbf{k}_{i,\|}, \hat{\mathbf{e}}_i) = \alpha^{(m)} \left\| \mathbf{J}^{(m)}(\mathbf{k}_{i,\|}, \mathbf{k}_{s,\|}) \hat{\mathbf{e}}_i \right\|^2, \quad (3)$$

where $\mathbf{J}^{(m)}(\mathbf{k}_{i,\|}, \mathbf{k}_{s,\|})$ is the Jones matrix of a single artificial meta-atom of the $m^{\text{th}}$ metasurface layer. This matrix contains the scattering coefficients between the incident and scattered planewaves in the basis formed by their two possible orthogonal linear polarizations [9]. It can be computed for particles of arbitrary shape embedded in layered media using virtually any Maxwell solver, combined with a near-to-far-field transformation. Fig. 3 illustrates two configurations involving a single metasurface: one placed atop a layered substrate, and another embedded within an Al$_2$O$_3$ layer on a Si substrate. Both are illuminated by the same normally incident plane wave with electric field $\mathbf{E}_i(\mathbf{k}_i, z)$. This excitation generates a background field $\mathbf{E}_b(\mathbf{k}_i, z)$, which forms a standing wave due to reflections from the substrate. The background field is computed under the assumption that all inclusions are absent; see Suppl. Note 2.4 for details.

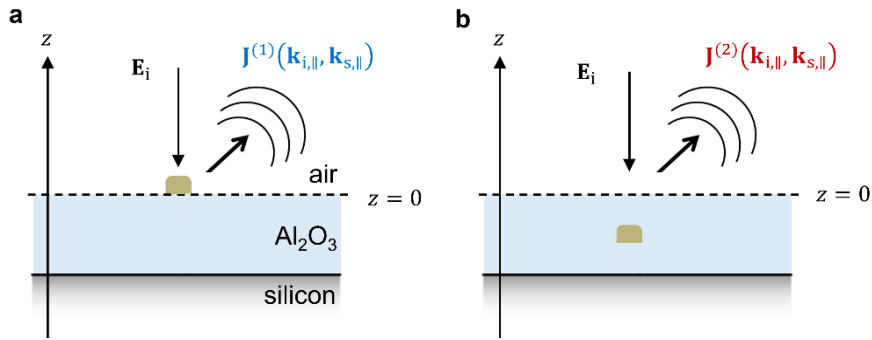

**Figure 3. Illustration of the computation of Jones matrices from a single meta-atom, a. laying of the layered substrate or b. buried in an Al$_2$O$_3$ layer**. Both situations result in two different background electric field $\mathbf{E}_b(\mathbf{k}_i, z)$ (generally a standing wave). The phase origin is unimportant for the model since we perform an incoherent sum. Note that the Al$_2$O$_3$ layer thickness varies with the number of metasurfaces; it is 190 nm for MS-1, 380nm for MS-2 and 570 nm for MS-3.

The correction factor—previously shown to be effective at moderately high particle densities, $\rho \approx 10\ \mu\text{m}^{-2}$, at optical frequencies [47]—is not included in the present simplified analysis. Accordingly, the ratio $C^{(m)}/(\cos\theta_i \cos\theta_s)$ in Eq. (2) is set to 1. However, we introduce a different correction factor, $\alpha^{(m)}$, in Eq. (3) to modify the form factor. This accounts for the fact that the $m^{th}$ metasurface layer is excited not by the background field (as in [9]), but by the locally averaged coherent field $\mathbf{E}_{\text{coh}}(\mathbf{k}_i, z)$. This coherent field is the mean-field solution predicted by the transfer matrix method used to model the specular component of light.

Specifically, for each metasurface layer, the excitation field is approximated as the average of the coherent field evaluated just above and just below the layer. This field incorporates both the original incident wave and the mean field scattered by all other layers. With this refinement, the model better accounts for the influence of multiple metasurfaces on the excitation of any individual layer.

Multiple scattering in the formation of the excitation field is incorporated through the correction factor $\alpha^{(m)}$, defined as the ratio between the coherent intensity $\left|\mathbf{E}_{\text{coh}}(\mathbf{k}_i, z^{(m)})\right|^2$ and the background intensity $\left|\mathbf{E}_b(\mathbf{k}_i, z^{(m)})\right|^2$, both evaluated at the position $z^{(m)}$ of the corresponding metasurface. Note that this ratio does not account for polarization conversion effects, which are negligible under normal incidence and for nanodisc geometries.

The proposed diffuse model (Eqs. 1–3) provides valuable physical insight by disentangling the distinct roles of the intrinsic scattering pattern of an isolated AuNP and its driving field $\mathbf{E}_{\text{coh}}$. Scattering from a metasurface is expected to peak at wavelengths where the driving field is strongest. Consequently, inter-particle interactions within the stack must modify the local excitation field in a way that enhances scattering at specific wavelengths [48].

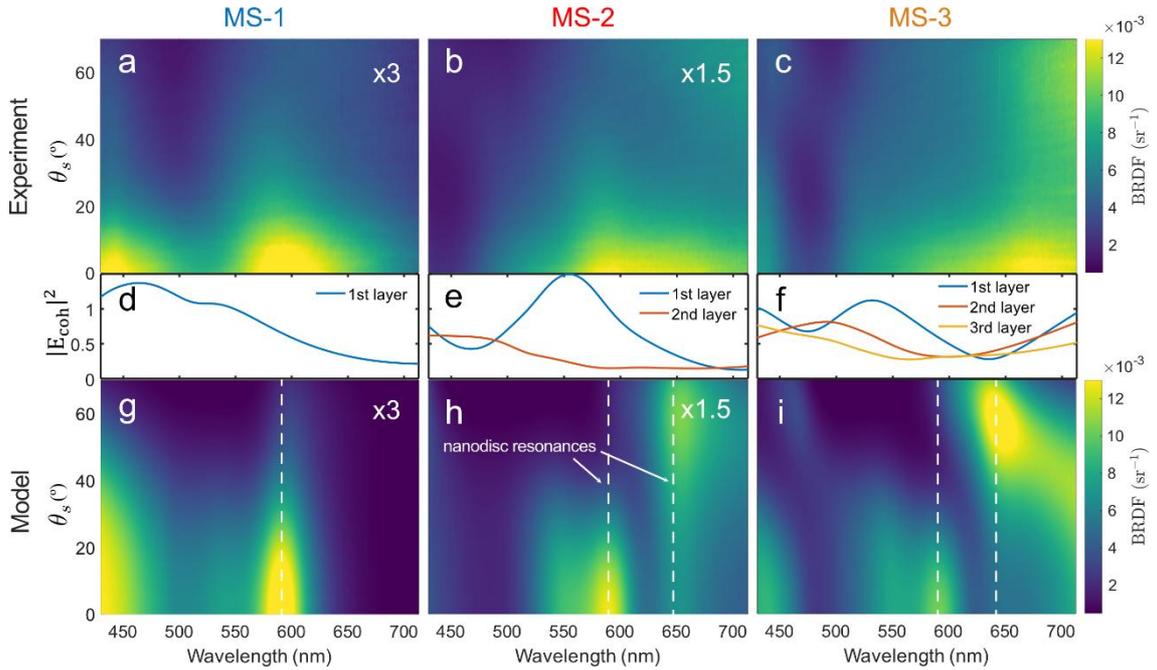

**Figure 4. Bidirectional reflection distribution function (BRDF) of cascading disordered metasurfaces. a-c.** Experimental diffuse BRDF map of single- (MS-1), double- (MS-2) and triple-layered (MS-3) metasurface stacks as a function of the viewing polar angle ($\theta_s$) and wavelength at normal incidence. Due to the $\theta_s \to -\theta_s$ symmetry at normal incidence, only measurements for positive angles are shown. BRDF intensities of MS-1 and MS-2 are multiplied by x3 and x1.5 respectively to allow easier comparison with MS-3. **d-f.** The coherent driving field of disordered metasurfaces in sample MS-1, MS-2 and MS-3. **g-i.** The BRDF maps of MS-1, MS-2 and MS-3 predicted using Eqs. (1-3) for normal incidence. The densities of disordered metasurfaces are 70 μm$^{-2}$ for MS-1, 80 μm$^{-2}$ for MS-2 and 90 μm$^{-2}$

for MS-3. Vertical dashed lines indicate resonance wavelengths of the nanodiscs, either laying on (at around 580 nm) or embedded (at around 650 nm) in an $Al_2O_3$ spacer layer.

**Comparison model vs experiment.**

Fig. 4a-c shows measured BRDF intensity maps as a function of wavelength and $\theta_s$ of our three samples, MS-1, MS-2 and MS-3, at normal incidence. The BRDFs are recorded with a home-made spectrogoniometre setup calibrated with a Lambertian diffusor using a supercontinuum laser as a source (See Methods and Suppl. Note 4). The diffuse reflection of MS-1 shows two peaks of similar intensity, with one blue peak centered around 440 nm, and another red peak centered around 580 nm. A broad peak in the red wavelengths (centered at 600 nm) emerges in the BRDF map of MS-2 while the blue peak in MS-1 disappears. Furthermore, for MS-3, a secondary peak seems to emerge in the red at large scattering angles. We additionally observe that the dominant peak of MS-3 at 670 nm is slightly red shifted compared to that of MS-2.

Fig. 4d-f presents the squared magnitude of the coherent electric field, $\left|\mathbf{E}_{\text{coh}}(\mathbf{k}_i, z^{(m)})\right|^2$, at the metasurface coordinates, $z^{(m)}$, computed from the specular model. All these fields are normalized by the incident electric field and exhibit various spectral profiles, mostly because of the different interferences for various $z^{(m)}$ values for MS-1, MS-2 and MS-3. Note the presence of vertical dashed lines in Fig. 4g-i which correspond to the resonance wavelengths of the nanodiscs, either laying (at around 580 nm) on or embedded (at around 650 nm) in an $Al_2O_3$ spacer layer (see Suppl. Note 2).

Fig. 4g-i shows the BRDF intensity maps computed from Eqs. (1-3). Again, the AuNPs are modeled as nanodiscs, with diameters of 60 nm for MS-1, 52 nm for MS-2, and 42 nm for MS-3. The height of all nanodiscs is 35 nm. The densities of AuNPs are estimated from the SEMs (see Suppl. Note 3), which is 70 $\mu m^{-2}$ for MS-1, 80 $\mu m^{-2}$ for MS-2 and 90 $\mu m^{-2}$ for MS-3. The model qualitatively predicts most of the observed features. It attributes the unique blue peak around 450 nm in MS-1 (Fig. 4a) to a maximum of the coherent excitation intensity $\left|\mathbf{E}_{\text{coh}}(\mathbf{k}_i, z^{(1)})\right|^2$. For MS-2, the maximum of the coherent excitation intensity for the top layer (blue curve) is slightly blue shifted compared to the resonance wavelength of the nanodisc in that layer, and the model predicts a diffuse intensity that spans from 600 to 650 nm, with two separated peaks. In MS-3, the intense peak between 650 nm and 720 nm, alongside weaker diffuse intensities below 450 nm and between 500 nm and 600 nm, corresponds with the observed local field peaks. In terms of scale, note the multiplicative numbers ×3 and ×1.5 for MS-1 and MS-2, respectively. More importantly, note that the quantitative agreement between the model predictions and the experimental results with the scaled BRDF values spanning from 1 to $12 \times 10^{-3}$ $sr^{-1}$.

There are also notable deviations between the model predictions and experimental results. For instance, the model predicts narrower resonance linewidths, which can likely be attributed to the polydispersity in the size and shape of the Au nanodiscs.

More intriguing is the model failure to accurately capture the angular dependence of the BRDF, particularly for sample MS-3. The model predicts a strong peak at large scattering angles around 650 nm, whereas the experimental data show a broader, yet more intense peak mostly at small scattering angles. The peak at large angles in the model arises from the form factor of individual nanodiscs. Due to the high reflectivity of the Si substrate, the dipolar resonances of the nanodiscs hybridize to form Fabry–Pérot-like modes, which can radiate at oblique angles $\theta_s$ when the distance between the nanodiscs and their mirror images in the substrate meets specific conditions [23].

There remains an additional puzzle as to why this phenomenon is not clearly observed in the measurements. It is known that at high surface densities, the diffuse component of the BRDF diminishes in favor of a stronger specular contribution, implying that the radiation cone—typically broad and Lambertian for small NPs in dense media—becomes narrower at large densities. This behavior is primarily due to multiple scattering, a mechanism largely neglected in the current model. Additionally, another source of discrepancy is our assumption that the nanodisc positions in different metasurface layers are statistically independent. In practice, some residual conformity likely exists due to the fabrication process and the interference term highlighted in Suppl. Note S2.1 (Eqs. S2.4-S2.5) might impact the directionality of the radiation.

Nonetheless, we believe that our model—which rigorously accounts for the $z$-variation of the excitation field and the layer-resolved scattering profile—has the potential to accurately and quantitatively describe diffuse reflection in metasurface stacks at moderately high surface densities (typically $\rho < 10$ μm$^{-2}$).

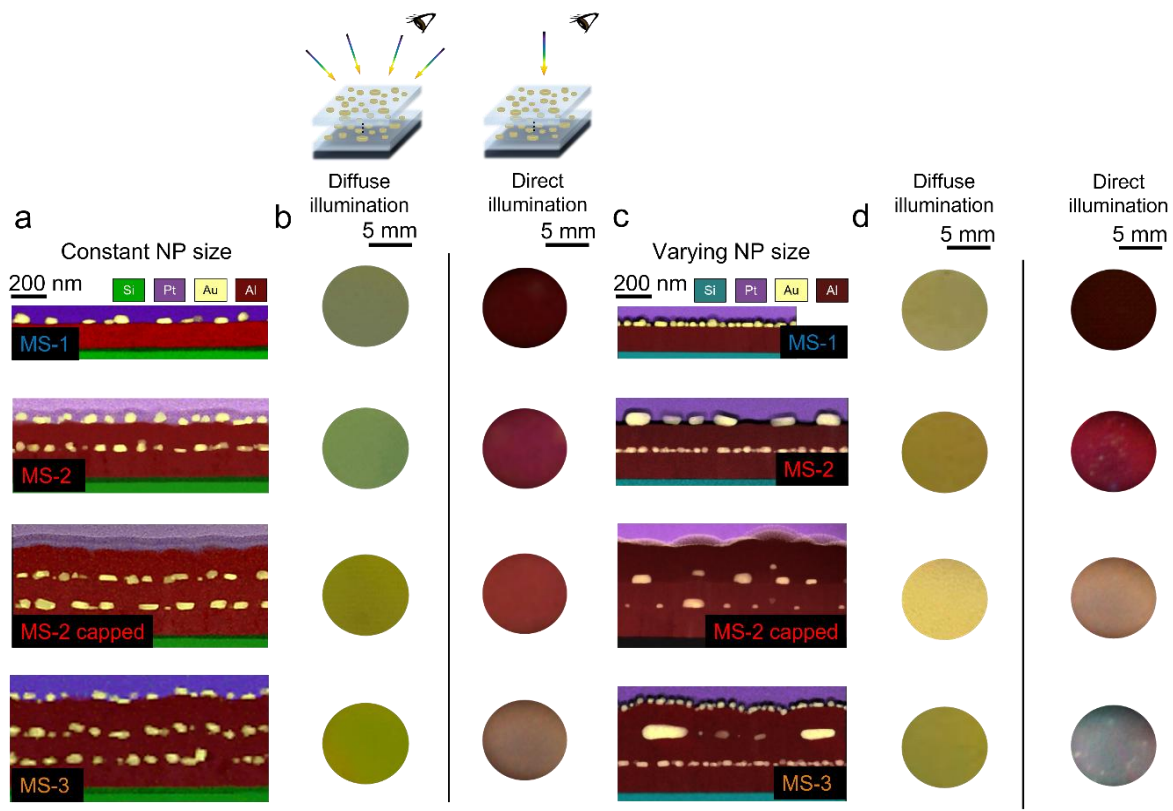

**Figure 5. Distinct visual appearance of cascading disordered metasurfaces under different illumination condition.** EDS-TEM images of cascading metasurfaces consisting of AuNPs of **a.** constant (50 nm) and **c.** varying (30-80 nm) average size. Inset: Colors stand for different materials. **b, d.** Photographs of the samples under (left) indirect illumination (room light) and (right) under a collimated solar simulator beam at normal incidence. All photos taken at an angle of 45°.

**Application to chromo-encryption.**

A unique characteristic of stacked disordered metasurfaces is that the interaction between layers modifies $\mathbf{E}_{coh}$ experienced by each constituent metasurface (Fig. 4d-f). Consequently, the overall diffuse scattering spectrum, and thus the perceived diffuse color, can be readily tuned by altering the number of metasurface layers (Fig. 4g-i). This principle can be the basis for chromo-encryption, a technique encoding information through the controllable color appearance of a surface [49]. Metasurface encoders used to encrypt information in telecommunication fiber networks [50] as well as in holographic and speckle-based architectures using wavelengths ranging from visible to THz regime [20,51,52] and various optical elements (polarisers, lenses, cameras) for information decryption.

To demonstrate this concept, we fabricated two sets of samples incorporating cascaded disordered metasurface layers composed of AuNPs separated by 150 nm $Al_2O_3$ spacers (TEM images in Figs. 5a, c).

- **Set 1 (Constant Nanoparticle Size):** These samples (MS-1, MS-2, MS-2 capped, MS-3) featured disordered metasurface layers of the same density, utilizing AuNPs of a consistent average size (50 nm), but varied the number of layers or included a capping $Al_2O_3$ layer (Fig. 5a).
- **Set 2 (Variable Nanoparticle Size):** In this set (MS-1, MS-2, MS-2 capped, MS-3), we introduced an additional tuning parameter by varying the average AuNP size from 30 nm to 80 nm across the different samples, alongside varying layer configurations (Fig. 5c).

The effectiveness of this approach becomes evident when comparing the samples' appearance under different lighting conditions (Fig. 4b, d). Under standard indirect room lighting, the samples in the first set appeared largely indistinguishable, sharing a similar greenish color (Fig. 4b, left). However, upon illumination with a direct, collimated white light source (solar simulator at normal incidence), their appearance transformed dramatically when viewed diffusely (camera at 45°). In Set 1, a clear color progression emerged: faint dark red (MS-1), vibrant red (MS-2), vermilion (MS-2 capped), and brown (MS-3) (Fig. 4b, right). According to Eq. (3), the diffuse light color is determined by the product of coherent driving field and the scattering diagram of single meta-atoms. Varying the averaged size of meta-atoms offers additional control over the diffuse light spectrum. Indeed, Set 2 demonstrated even broader color control due to the combined effects of layering and NP resonance tuning, exhibiting diffuse colors ranging from black, through red and brown, to white across the different samples under direct illumination (Fig. 4d, right).

## CONCLUSIONS

Using a scalable and lithography-free fabrication and a theoretical framework, we assemble and study the optical properties of cascading disordered metasurfaces. Engaging ALD with Angstrom scale thickness precision and sputtering, we achieve three-layered disordered metasurface stack with an overall sub-micron thickness. We perform angle-resolved optical experiments to measure the specular and diffuse reflection of the resulted cascading metasurfaces.

An important contribution of this work is the derivation of semi-analytical models that predict the properties of the diffuse light, even for large incidence angles. It highlights that the color of the incoherent light is driven by not only the individual nanoparticle resonances, but also by the color of the coherent light at the metasurface locations, revealing how the diffuse color can be tuned by varying the number of metasurface layers and their separation distances.

We apply this understanding to design a chromo-encryption device composed of three metasurface layers. The stacks exhibit similar appearances under diffuse illumination but display strikingly distinct colors under direct light illumination. This unique illumination-dependent characteristic holds significant promise for applications in chromo-encryption, such as visual authentication of documents or banknotes, which can be verified by the naked eye without requiring sophisticated optical apparatus. Further investigations using a systematic study of nanoparticle morphologies, densities, materials will provide more information about the dominant optical processes at this sub-wavelength scale.

Insights from such cascading disordered media are extremely valuable for designing future nanophotonic platforms inspired by traditional optical systems such as stacked lenses in cameras and objectives, where layering is a natural and potent strategy [29,53] to enhance performance and unlock new functionalities, all while maintaining miniaturization and small footprint.

**METHODS**

Sample preparation: ALD/sputtering deposition and film annealing.

ALD was used to deposit $Al_2O_3$ on a polished side of a 4-inch Si (100) wafer. The deposition was carried out in a custom-built deposition module (SC-1, Swiss Cluster) at a temperature of 225 °C using Trimethylaluminium (TMA) and $H_2O$ precursors. This was followed by the deposition of a 15 nm Au layer using direct current magnetron sputtering in a vacuum chamber with a $5.0 \times 10^{-7}$ Torr base pressure integrated within the same deposition chamber. For deposition of the Au film, we used $5 \times 10^{-3}$ mbar working pressure and 40 W sputtering power at a deposition rate of 0.45 nm/s. Further the sample was annealed in a rapid thermal annealing furnace (RTA: ULVAC-RIKO MILA-5000-P-N). Annealing experiments were performed at 850 °C for 180 min. To avoid contamination, the samples were placed on a bare sapphire substrate, which then was placed on a quartz holder in the furnace. The metal layer was transformed into uniformly distributed AuNPs by solid-state dewetting. By repeating the above-mentioned steps, we created two- (MS-2) and three-layered (MS-3) MS stacks.

Optical measurements.

All optical measurements were performed with an in-house goniospectrophotometric setup devoted to the appearance of metasurfaces [1], which used a supercontinuum source and a solar simulator for the illumination and a Canon EOS1000D and a spectrometer for the photo and spectra acquisition, respectively. Two concentric DC motor rotation stages (Newport, URS75 and URS150) and a vertical arm controlled the incident and scattering (viewing) angles, $\theta_i$ and $\theta$, accordingly. Briefly, for the diffuse light BRDF measurements, we used an unpolarized centimeter-scale expanded supercontinuum laser beam (Leukos, Rock 400). A short-pass filter (Schott, KG-1) to keep only the visible spectrum range. The backscattered light was collected slightly above the plane of incidence by a 1 mm core diameter optical fiber connected to a spectrometer (Ocean Insight, HDX). The incident laser radiant flux was measured with the same setup and with the fiber detector facing the focused laser beam.

SEM, TEM and HAADF characterisation.

Scanning electron microscope (SEM, Tescan Lyra), HAADF, EDS integrated into a TEM (Thermo Scientific Titan 200 G3 at 200 KeV) are used for the nanoscale structural characterization of the disordered MS stacks.

**AUTHOR INFORMATION**

* Those authors contributed equally.

**ACKNOWLEDGMENTS**

A. X. acknowledges support by the NTNU Rector AVIT fund (grant number, 949024104) and NTNU NanoLab via the Research Council of Norway (RCN) support to the Norwegian Micro- and Nano-Fabrication Facility, NorFab (grant number 295864). M.C. acknowledges support from the China Scholarship Council (grant number 202002527052).